# THE UNFOLDING OF THE SPECTRA OF LOW ENERGY GALACTIC COSMIC RAY H AND HE NUCLEI AS THE VOYAGER 1 SPACECRAFT EXITS THE REGION OF HELIOSPHERIC MODULATION

W.R. Webber[1], P.R. Higbie[2] and F.B. McDonald[3+]

1. New Mexico State University, Department of Astronomy, Las Cruces, NM 88003, USA

2. New Mexico State University, Physics Department, Las Cruces, NM 88003, USA

3. University of Maryland, Institute of Physical Science and Technology, College Park, MD 20742, USA

+ Deceased August 31, 2012

# ABSTRACT

This paper describes the unfolding of the solar modulated galactic cosmic ray H and He nuclei spectra beyond ~105 AU in the heliosheath. Between 2008.0 and 2012.3 when Voyager 1 went from about 105 to 120.5 AU the spectral intensities of these two components between about 30 and 500 MeV/nuc unfolded (increased) in a manner consistent with an average modulation potential decrease ~5 MV per AU as described by a Parker like cosmic ray transport in the heliosphere where the overall modulation is described by a modulation potential in MV. Between 120.5 and 121.7 AU, however, as a result of two sudden intensity increases starting on May 8[th] and August 25[th], 2012, this modulation potential decreased by ~80 MV and spectra resembling possible local interstellar spectra for H and He were revealed. Considering these spectra to be the local interstellar spectra would imply that almost 1/3 of the total modulation potential of about 270 MV required to explain the spectra of these components observed at the Earth must occur in just a 1 AU radial interval in the outer heliosheath. As a result about ~80% of the total modulation potential observed at the Earth at this time occurs in the heliosheath itself. The remaining 20% of the total modulation occurs inside the heliospheric termination shock. The details of these intensity changes and their description by a simple modulation model are discussed.

**Introduction**

In late 2004 Voyager 1 crossed the heliospheric termination shock (HTS) at a distance of 94 AU from the Sun at a latitude of 34°N (Stone, et al., 2005). In analogy to what was observed at V2 when it crossed the HTS later in 2007, the radial solar wind speed at V1 most likely decreased from an average ~400 km·s$^{-1}$ to ~130 km·s$^{-1}$ (Richardson, et al., 2008) and then later to a value close to zero (Krimigis, et al., 2011), and the magnetic field became larger and more turbulent (Burlaga, et al., 2005). The cosmic ray diffusion coefficient in this field therefore became much smaller after the shock crossing.

As a result of these parameter changes, the process by which galactic cosmic rays (GCR) move inward in the heliosphere in this region beyond the HTS could be quite different than the usual Parker process (1958) which describes cosmic ray modulation inside the HTS. however it may still be characterized as a modulation potential. The Parker process assumes an expanding solar wind with a Betatron like deceleration process, which can be described as a potential difference, along with convection and diffusion.

The intensities of GCR nuclei and electrons are indeed observed to increase substantially beyond the HTS (e.g., McDonald, et al., 2006, 2012). For electrons from ~6-100 MeV this increase, up to 2012.7 when V1 is at 121.7 AU, is a factor of at least 100 in the heliosheath. This increase occurs over a distance of ~27.7 AU between the HTS crossing distance and 121.7 AU. This compares with a factor ~2 or less increase of these electrons between the Earth and the HTS over a distance ~90 AU.

For cosmic ray nuclei the unfolding of the H and He nuclei spectra is more modest, but at an energy of 145 MeV for protons the intensity increase beyond the HTS is nearly a factor ~10 and for He nuclei at this same energy, the factor is ~5. For these same particles between the Earth and 72 AU, corresponding to the minima in the solar modulation in 1977 and 1998, the radial increase observed by Voyagers 1 and 2 is a factor of about 3 for protons and ~2 for He nuclei. Thus most of the solar modulation of these GCR components at these lower energies also appears to be taking place in the region beyond the HTS.

In the 1$^{st}$ 3 years after the HTS crossing (about 11 AU in outward motion for V1) the intensities of all components generally increased and the apparent radial intensity gradients, especially for electrons, were relatively large and variable. These gradients then become more regular beyond about 105 AU. At distances of 111 and 116 AU, sudden ($\leq$ 1-26 day solar rotation period) intensity changes and radial gradient changes are observed for electrons (Webber, et al., 2012). These changes appear to be caused by the passage of V1 into different regions (structures or sectors) at these locations which are at distances of ~17 and 22 AU beyond the HTS crossing distance (e.g., Florinski, et al., 2011). Throughout this time period the energy spectra of H and He nuclei have unfolded in a manner consistent with that to be expected in a simple "Parker like" modulation model where the potential difference between the Voyager location and that representing the cosmic ray source spectra is decreasing (Webber, 2012). On average the potential difference associated with this modulation appears to decrease by about 20-25 MV per year from 2008 to 2012 as Voyager goes from ~105 to 120 AU.

It is the object of this paper to chronicle this spectral unfolding for H and He nuclei, extending the measurements to beyond 120 AU where the changes are even more dramatic than those inside 120 AU, and to interpret all of these changes in the simplest available modulation model, albeit that the conditions, particularly in the outer heliosheath being traversed by V1, are most unusual.

## A Description of the Data

In Figure 1 we show the radial intensity increases for H and He nuclei for various energies between about 2006.6 and the end of 2012 when V1 is between 100 and 123 AU. The data represents 6 month averages until 2011 when the time intervals become finer; eventually decreasing to 26 day intervals in 2012. This is to show, in particular, the two significant intensity increases of both nuclei and electrons in 2012. However, in the time from about 2007 to late 2011 the apparent radial intensity gradient is, on average, almost constant for each of the individual energies/charges that are plotted. This average gradient for each curve in %/AU is shown in the Figure.

These last two increases and their relative magnitude are shown in more temporal detail in Figure 2 for nuclei >200 MeV. The total increase of the high energy nuclei is ~31% with 14% of this increase occurring in the 1[st] step. A stepwise increase of this magnitude is unprecedented.

The data in Figures 3 and 4 is in the form of energy spectra for H and He nuclei obtained from the HET telescopes on the Voyager CRS experiment (Stone, et al., 1977), see also Webber and Higbie, 2009, for details on how the spectra are obtained. Also included in the figures are curves calculated for a modulation model to be described later with parameters chosen to match the data. The study of the spectral details for electrons and heavier nuclei in this time period will be deferred to later papers but are also characteristic of a large sudden change in the modulation potential. The figures 3 and 4 for the H and He nuclei are matched in intensity and energy scale so that the figures can be simply over-laid to compare the relative intensity changes. The figures show the intensities measured at the following times: 2008.0 (105.0 AU), 2010.0 (112.2 AU), 2012.0 (119.3 AU) and a period after 2012.7 (122.0 AU). The data at 2012.0 and after 2012.7 comes from the paper by Stone, et al., 2013. Each spectral point for 2008.0 and 2010.0 is a one year average centered on the dates indicated so that short term changes are smoothed out. Also shown in the two figures, in addition to the V1 spectra after 2008.0, are the H and He nuclei spectra measured in 1998-99 when V1 was at 72 AU, still inside the HTS, but at a time when intensities were at a maximum in the 11 year solar cycle (Webber, McDonald and Lukasiak, 2003).

For comparison to this Voyager data we show the spectra reported for H and He nuclei by the PAMELA experiment near the Earth at the end of 2009 (Adriani, et al, 2013) at the time of the highest intensity maximum for cosmic rays of modern times (McDonald, et al., 2010; Mewaldt, et al., 2010).

The Voyager spectra presented here are background corrected when necessary. For protons and He nuclei a correction is made for lower energy anomalous particles. At 100 MeV this correction at 2012.0 and earlier times is ~50% of the total measured intensity for protons. At this same energy this correction is ~15% for He nuclei. These low energy anomalous particle spectra have exponents between ~-2.5 to -3.5 at ~100 MeV and above so that this correction rapidly becomes less at higher energies. At lower energies this limits the energy to which the GCR proton spectrum can be derived (taken to be ~50% background) at 2012.0 and earlier to about 80 MeV. For He this minimum energy is ~60 MeV/nuc.

The proton spectra in Figure 3 therefore cover the energy range from 80-350 MeV and the Helium spectra in Figure 4 from 60-630 MeV/nuc.

In Figure 3 for protons the total unfolding (increase) from 2008.0 to the time period after the final increase starting at 2012.35 is a factor ~3.61 at 145 MeV; at 310 MeV this unfolding factor is 2.16 and this factor is systematically decreasing with increasing energy. For He nuclei in Figure 4 this total unfolding factor at 62 MeV/nuc is 5.18; at 145 MeV/nuc it is 2.73 and at 310 MeV/nuc this factor is = 1.55, again systematically decreasing with increasing energy but now the fractional increase is much smaller than for protons at the same energy.

Although the intensities in 1998-99 at 72 AU inside the HTS measured by V1 are at an intensity maximum in the 11 year solar modulation cycle they are still well below what was observed for the same species later in 2008.0 at 105.0 AU in the heliosheath at a time when the overall solar modulation potential was much greater. The intensity increase between 1998 at 72 AU and 2008.0 at 105 AU at V1 is a factor ~1.45 for protons at 145 MeV and ~1.28 for He at the same energy/nuc. And since 2008.0 is a time when the solar modulation effects at the Earth were actually greater than at 1998-99, the intensities at 105.0 AU in the heliosheath might have been expected to be less than in 1998-99 at 72 AU due to the increased solar modulation at that time. This underscores the rapid increase of intensities in the heliosheath.

Of special interest in this paper are the spectral changes taking place at 2012.35 and 2012.65. This is an overall time period when the radial distance to V1 changes by only 1.1 AU from 120.6 to 121.7 AU. At ~145 MeV, the lowest energy for which the anomalous cosmic ray (ACR) background can accurately be subtracted from the total proton spectrum at 2012.0, the spectral unfolding factor is 1.90 for this short time period, which is about the same as the total intensity increase for the 4 years between 2008.0 and 2012.35 when V1 moved outward 15.5 AU. At 310 MeV the unfolding factor for this same short time period is 1.50 for protons, again about the same as the increase between 2008.0 and 2012.0.

For He nuclei at 62 MeV/nuc the unfolding factor for the time interval between 2012.35 and 2012.65 and between 2008.0 and 2012.0 are 2.28 and 2.25 respectively; at 145 MeV/nuc for He they are 1.64 and 1.55 respectively; at 310 MeV/nuc they are 1.28 and 1.22 respectively. Again these ratios are similar for both time periods covering ~1.1 AU and 15.5 AU respectively of outward movement by Voyager, indicting a similar energy/rigidity dependence of the intensity changes in both time periods.

To summarize, for H and He nuclei above about 60 MeV/nuc, the intensity changes defining the overall modulation are similar in the two time periods, the 1[st] time period covering the ~1 AU thick region at 121 AU and the second time interval covering the outer 15.5 AU of the heliosheath between 105-120.5 AU.

## Comparison of Voyager Data with Data at the Earth

Here we compare here the Voyager intensities with the intensities for protons and Helium nuclei that have been reported from the PAMELA experiment. This experiment has been operating over the time period from 2006 to 2012 and therefore covers the time of intensity maximum at the Earth in late 2009.  The PAMELA data, (Adriani, et al., 2013) is also shown in Figure 3 for protons and Figure 4 for He nuclei.

For both protons and Helium nuclei it is seen that the Voyager and PAMELA data sets cover similar energy and time regimes.  The intensities for both protons and He nuclei measured by PAMELA at the time of the intensity maximum in solar cycle #24 in 2009 at the Earth are comparable to but slightly less than those measured by Voyager 1 at a distance of 72 AU at the time of maximum intensity in 1998 in solar cycle #23, 11 years earlier.  This similarity of intensities at 1 and 72 AU suggests; (1) There is a rather weak amount of overall modulation throughout the inner heliosphere; (2) The uniqueness of the high intensities of these nuclei observed at the Earth in 2009 (e.g., McDonald, et al., 2010; Mewaldt, et al., 2010).

These high intensities near the Earth at 2009 are characterized by the extremely low modulation potentials ~250-270 MV that are required to reproduce the intensities and spectra of H, He and heavier nuclei that are observed at this time (e.g., Wiedenbeck, et al., 2012).

**Cosmic Ray Transport in the Heliosphere**

Here we consider a spherically symmetric quasisteady state two radial zone (or hybrid) no-drift transport model for cosmic rays in the heliosphere.  This model has been previously used by us on several occasions to describe the modulation of protons, Helium nuclei and Carbon nuclei measured by V1 and V2 between ~2005 and 2011 when V1 and V2 were both inside and outside the HTS at distances of between 70 and 115 AU from the Sun (Webber, et al., 2011a, 2011b and 2012).  While this simplified model, which does not include drifts, obviously cannot fit all types of observations it does provide a useful insight into the relative inner heliospheric/outer heliospheric modulation and helps to determine which aspects of this modulation need more sophisticated models for their explanation.  The numerical model was originally provided to us by Moraal (2003, private communication) and is similar to the model described originally in Reinecke, Moraal and McDonald, 1993, and in Caballero-Lopez and Moraal, 2004, and also similar to the spherically symmetric transport model described by Jokipii, Kota and Merenyi, 1993 which does include drifts (Figure 3 of that paper).  The basic transport equation is (Gleeson and Urch, 1971);

$$\frac{\partial f}{\partial t} + \nabla \cdot (CVf - K \cdot \nabla f) + \frac{1}{3p^2} \frac{\partial}{\partial p} (p^2 V \cdot \nabla j) - Q$$

Here f is the cosmic ray distribution function, p is momentum, V is the solar wind velocity, K(r,p,t) is the diffusion tensor, Q is a source term and C is the so called Compton-Getting coefficient. The final term on the left includes energy loss.

For spherical symmetry (and considering latitude effects to be unimportant for this calculation) the diffusion tensor becomes a single radial coefficient $K_{rr}$. We assume that this coefficient is separable in the form $K_r(r,P) = \beta\ K_1(P)\ K_2\ (r)$, where the rigidity part, $K_1(P) \equiv K1$ and radial part, $K_2(r) \equiv K2$. The rigidity dependence of $K_1(P)$ is assumed to be ~P above a low rigidity limit $P_B$. The units of the coefficient $K_{rr}$ are in terms of the solar wind speed V = 4 x $10^2$ km·s$^{-1}$ and the distance 1 AU=1.5 x $10^8$ Km, so $K_{rr}$ = 6 x $10^{20}$ cm$^2$·s$^{-1}$ when K1 = 1.0.

Based on our earlier modulation studies at the Earth and V2 and V1 using protons, Helium nuclei and Carbon nuclei (Webber, et al., 2011a, 2011b, 2012) we consider a distinct two zone heliosphere (as was done by Jokipii, Kota and Merenyi, 1993). In this case the inner zone extends out to 94 AU, the average distance to the HTS. In this inner region V=400 km·s$^{-1}$ and the diffusion parameters K1 and K2 are determined in our approach by a fit to the cosmic ray data being compared.

The outer zone extends from the average HTS distance of about 94 AU to ~122 AU, the approximate distance to the equivalent "outer modulation boundary". This region is essentially the heliosheath. In this region V is taken to be 130 km·s$^{-1}$ (from V2 measurements, Richardson, et al., 2008) and the diffusion parameters are K1H and K2H, which are greatly different from those in the inner heliosphere are again determined by the fit to the cosmic ray intensity changes observed at V1. The location of this outer boundary and the LIS spectra are important in this calculation.

For the LIS H and He spectra we are fortunate that during the writing of this article, V1 appears to have exited the main solar modulating region as a result of the two sudden and large increases at 2012.35 and 2012.65 mentioned earlier. Also at 2012.65, as a result of the disappearance of energetic ions ~1-80 MeV (ACR), V1 also appears to have exited the confinement region for these low energy particles within the heliosheath. The intensity decrease of these ACR is so large (over a factor ~100) and so rapid (a few days) that the entire and much weaker GCR H and He low energy spectra are suddenly revealed, perhaps down to energies of a few MeV. Furthermore, the higher energy GCR nuclei (~100-200 MeV and above) increased by a factor of up to 1.5 in just a few days at the same time the ACR decreased.

Thus for the LIS spectra needed in this paper we will assume that the intensities measured by Stone, et al., 2013, at V1 between about 2012.75 and 2013.0 for H and He nuclei are indeed the LIS ones. These observed spectra can be approximated to an accuracy ~few % above a few MeV by:

$$H \text{ FLIS} = (20.2/T^{2.70})/(1+(6.75/T^{1.22})+(0.50/T^{2.80})+(0.0002/T^{4.42}))$$

$$He \text{ FLIS} = (0.94/T^{2.70})/(1+(4.14/T^{1.09})+(0.30/T^{2.79})+(0.00012/T^{4.26}))$$

where T is in GeV/nuc.

To illustrate the diffusion coefficient that is used in the calculation in the two zones of the heliosphere, we show Figure 5 which presents the values and rigidity dependence of this parameter. At rigidities greater than ~300 MV, which includes most of the GCR nuclei, this coefficient is taken to be ~$P^{1.0}$ and has a value in the inner heliosphere, K1=200 at 1 GV (=200x6x$10^{20}$)= 1.2x$10^{23}$ cm$^2$·s$^{-1}$. In the heliosheath the value of K1, K1H, is taken to be =12 (16.5 times smaller) at 1 GV.

The calculated modulated spectra corresponding to successive modulation potential differences of 8 MV corresponding to the three 1 AU radial intervals between 119-122 AU are shown in Figures 3 and 4 respectively for H and He nuclei. The radial spacing is then changed to 7 AU (119, 112 and 105 AU) corresponding to total modulation potentials of 24, 80 and 136 MV with reference to the boundary taken to be 122 AU (in one year the V1 radial distance increases by 3.6 AU). The modulated spectra are also shown at the HTS at 95 AU (about 220 MV) and at the Earth (270 MV) for these same diffusion coefficients.

These calculated spectra give a good representation of the observed spectral unfolding at V1 between 2008.0 to 2012.0 for total modulation potentials of between 136 and 80 MV respectively for H and 160 and 110 MV respectively for He nuclei.

At 2012.0 the spectra of H and He would therefore require a further modulation potential decrease ~80 and ~110 MV respectively for H and He nuclei at an energy ~145 MeV/nuc in order to reach the FLIS representation of the LIS as observed by V1 later in 2012. This decrease in potential would need to occur over a distance ~1 AU.

**General Comments Regarding the Overall Modulation in the Heliosphere**

The modulated spectra we have presented in Figures 3 and 4 represent a very simple overall modulation picture at a single time and with radially independent diffusion coefficients in the two radial zones of the heliosphere. Although there is general good agreement with the data there are a number of specific differences between the predictions and observations that deserve further comment. But before doing this, we should note the following. The total modulation potential from the boundary at 122 AU to 1 AU is ~270 MV using these parameters. Using this value, we are able to fit the measured intensities of H and He nuclei at the Earth in late 2009 measured by PAMELA (Adriani, et al., 2013) at a time when this modulation potential reached its lowest values since accurate modern data have been available (McDonald, et al., 2010; Mewaldt, et al., 2010). This value is consistent with the total modulation potential as obtained from the Carbon spectrum measured by the CRIS-SIS experiment on ACE at this time which was ~250 MV (Wiedenbeck, et al., 2012).

Of the 270 MV modulation potential that is calculated at the Earth at this time, ~80 MV occurs in the two sudden steps near 121 AU. The total modulation in the rest of the heliosheath is ~5 MV/AU x 28 AU, the thickness of the heliosheath, giving a total of ~220 MV in the entire heliosheath. The remaining 50 MV occurs in the region from 1 AU out to the HTS at 95 AU. However, this still produces a significant radial gradient in the inner heliosphere region particularly at higher energies.

Note that these modulation potentials are for sunspot minimum conditions. Additional modulation occurring during the normal solar 11 year modulation cycles which increases this modulation potential is believed to occur mainly within the inner heliosphere, i.e., out to the HTS at approximately 95 AU.

Similar modulation fractions for the outer and inner heliosphere are described by Potgieter (2008).

**Modulation at Sunspot Minimum in Opposite Solar Polarity Cycles – the 1998-99 V1 data at 72 AU and Recent PAMELA Data in 2009**

Once the LIS spectrum is actually available the overall heliospheric modulation potential can be well determined by intensity measurements at the Earth. What is much less certain is the relative fraction of the modulation potential that occurs inside the HTS to that between the HTS and the modulation boundary. For example, in our model, at 2009 at the solar cycle minimum, this relative fraction of inner to total heliospheric modulation is calculated to be 0.18 (50/270) MV. In other words 18% of the modulation potential occurs inside the HTS.

In 1998-99 at the positive solar cycle #23 modulation minimum 11 years earlier, the intensities of 145 MeV H nuclei measured at V1 at 72 AU are somewhat higher than those measured by PAMELA at the Earth in 2009, but only by a factor ~1.30. In 1998, however, the ratio of H nuclei intensities between 72 and 1 AU is ~2.0 at this energy. The ratio of the LIS intensity to that at 1 AU at this time is ~10. So that from these measured intensity ratios at a single time, roughly 20% of the total modulation is occurring between 1 and 72 AU, compatible with our estimates above.

## The Spectral and Intensity Changes of H and He Nuclei Starting on May 8[th] and August 25[th], 2012

It is evident from Figure 1 of this paper, which describes the time history of intensities during the time period that V1 is passing through the outer heliosheath after about 2006.0, that the grand finale of GCR increases in 2012 occurred in two steps. These steps are shown in Figure 2. The 1[st] increase started on May 8, 2012, and took about 26 days to complete. It was followed by a plateau of almost constant intensity for GCR nuclei and electrons which lasted about 52 days. Then two sudden temporary increases in GCR of duration ~3 days and ~7 days, respectively, occurred with the onsets of the increases separated by ~16 days. This was followed by the final increase which started 13 days after the onset of the last of the two precursor increases. This final GCR increase was basically completed in ~7 days for the nuclei considered here and was the largest intensity increase of these GCR components since the launch of V1 35 years ago.

The most statistically accurate measure of the comparative increases during steps 1 and 2 comes from the PGH rate (H$\geq$200 MeV, e>12 MeV) shown in Figure 2. The increases in this rate are 13.3±0.7% for the 1[st] increase and 33.4±1.4% for the total increase.

Regarding the rigidity/energy dependence of the overall two step increase of GCR H and He nuclei, which is one of the most important aspects defining the mechanism of this episode of modulation, we show in Figure 6 a plot of the calculated intensity changes, ln $j_2/j_1$ vs. P, for H nuclei for a potential difference (E loss) of 80 MV in our model. This calculation is also shown for He nuclei in the same figure. This figure illustrates a $P^{-1.0}$ dependence of the modulation above ~1 GV where the H and He input spectra are taken to be ~$P^{-2.0}$. This dependence of ln $j_2/j_1$ vs. P is characteristic of our choice of the $P^{1.0}$ rigidity dependence of the diffusion coefficient. A turn up and splitting of the H and He intensity changes is calculated to occur at lower rigidities. This splitting is due to differences in the LIS rigidity spectra of H and He at lower rigidities and also to the different rigidities at which the intensity maximum occurs.

The actual observed intensity changes during this time period for H and He nuclei are shown in Figure 6 alongside the predictions. The agreement between predictions and observations for H nuclei is excellent over the entire rigidity range. The agreement for He nuclei is less good, however. A "splitting" between the magnitude of the H and He intensity changes is indeed observed at a fixed rigidity, but this splitting appears to be larger than indicated by the calculations.

From this comparison it appears that the final 2 step increase exhibited spectral intensity changes that were consistent with an E loss mechanism.

## Summary and Conclusions

The paper presents a summary of the unfolding of the energy spectra of H and He nuclei in the heliosheath as measured on V1. This spectral unfolding occurs between about 105 and 122 AU as V1 approaches the outer boundary of the main heliospheric modulating region. The observed spectral unfolding is well represented to first order by the spherically symmetric two zone modulation model used here, a "Parker" based diffusion-convection model. Based on this model the V1 data on H and He nuclei would suggest an energy loss potential difference that decreases by 20-25 MV/year from a value ~180 MV to 80 MV between 2008 and 2012 when V1 is between 105-120 AU in the heliosheath. Between 120.6 and 121.7 AU an additional modulation potential decrease of 80 MV is necessary to explain the two sudden increases of the intensity of lower energy H and He nuclei that occur in 2012. This potential difference in a 1 AU transition region amounts to ~1/3 of the total modulation potential of 270 MV between the Earth and 121.7 AU at this time.

Although this unfolding of H and He nuclei spectra in the heliosheath can be reasonably well fit using a Parker-like model with a potential energy loss, using the formulation that has been originally applied to the inner heliosphere, most of the total heliospheric modulation at this time is actually taking place well beyond the HTS which is located at about 94 AU. In the heliosheath the value of the diffusion coefficient required to produce this modulation is only about 1/15[th] of its average value inside the HTS.

This modulation even continues to take place in the region beyond ~112 AU where Krimigis, et al., 2011, have determined that the average solar wind speed (the convective and E loss terms in the Parker equation) approaches zero.

The large modulation taking place between 120.6 and 121.7 AU is difficult to understand physically although its overall effect is described quite well by the simple models we use here. For H nuclei the changes resemble those to be expected for a total potential loss ~80 MV taking place in two almost equal steps. For He nuclei this potential loss is slightly larger. For this energy loss to occur over ~1 AU in a Parker like modulation would require a diffusion coefficient (literally a diffusion barrier) ~10 times smaller than the already small average diffusion coefficient that we have used to describe the heliosheath spectral intensity unfolding between 105 and 120 AU.

As a result of the final unfolding between 120.6 and 121.7 AU, a spectrum of lower energy H and He nuclei is revealed down to perhaps 2 MeV. This spectrum is very similar to that estimated from simple diffusion models of cosmic ray transport in the galaxy (e.g., Ip and Axford, 1985; Putze, et al., 2011). Comparing the calculated spectra for Parker like modulation levels which use modulation potentials of only a few MV with the actual measured H and He spectra at these lower intensities restricts possible lower level modulation effect outside of the normal heliosphere to probably less than ~10 MV of a Parker-like modulation potential (see, however, Strauss, et al., 2013).

Other localized effects on the LIS spectra as described by Stone, et al., 2013, should be observable from the intensity gradient as V1 proceeds further beyond this boundary. New specific galactic propagation models extending down to a few MeV are needed to define the expected LIS spectrum in this here-to-for unexplored region of the spectrum.

<u>**Acknowledgements**</u>

We appreciate the support of JPL for the Voyager program and from our colleagues Ed Stone, Project Manager, and Alan Cummings on the West Coast and Nand Lal and Bryant Heikkila on the East Coast. The H and He data used here comes from the Stone, et al., 2013, Science paper and from internal web-sites maintained by Nand Lal and Bryant Heikkila.

## Figure Captions

**Figure 1:** Intensity changes of H and He nuclei for various energies between 2005 and 2012 when V1 is between ~100-123 AU in the heliosheath. The values of the average radial gradient between 2005.0 and 2012.0 in %/AU are shown next to the intensity lines. The data are 6 month averages up to 2012.0 and 26 day averages in 2012.

**Figure 2:** Five day running averages of the >200 MeV rate from 2011.8-2012.9 showing the relative magnitude of the two increases starting on May 8 and August 25.

**Figure 3:** Yearly average H nuclei intensities centered at 2008.0, 2010.0 2012.0 and also the time period from 2012.70 to 2013.0. The intensities measured at V1 in 1998-99 at the time of the previous modulation minimum are also shown as well as the PAMELA intensities in late 2009 at the Earth. Calculated spectra are shown for modulation levels in MV with reference to the FLIS=LIS. The shaded band shows the amount of modulation occurring in the inner heliosphere between the Earth and the HTS.

**Figure 4:** Same as Figure 2 but for He nuclei.

**Figure 5:** The values and rigidity dependence of the diffusion coefficients used in the inner and outer heliospheric modulating regions. Note the sudden break of the rigidity dependence from $\sim P^{1.0}$ to $\sim P^{-1.0}$, at 300 MV in the inner zone and at 150 MV in the outer zone. The rigidity dependence and magnitude of the diffusion coefficients appropriate to lower rigidity electrons that are shown as a shaded region in the Figure are quite different than those above a few hundred MV that apply to the nuclei being modeled here.

**Figure 6:** TOP - The fractional intensity change, $\ell n\ j_2/j_1$, vs. P (MV) for H and He nuclei calculated using our model parameters for a decrease in modulation potential from 80 MV to zero (solid lines). The observed intensity changes at V1 from before May 8[th] to after August 25, 2012 are shown as open circles, black for H nuclei between 79 and 570 MeV and red for He nuclei between 62 and 570 MeV/nuc. A fractional intensity change $\sim P^{-1.0}$ is also shown for reference.

BOTTOM – Average radial intensity gradients in %/AU between 100 and 120 AU for H (red) and He nuclei (black).

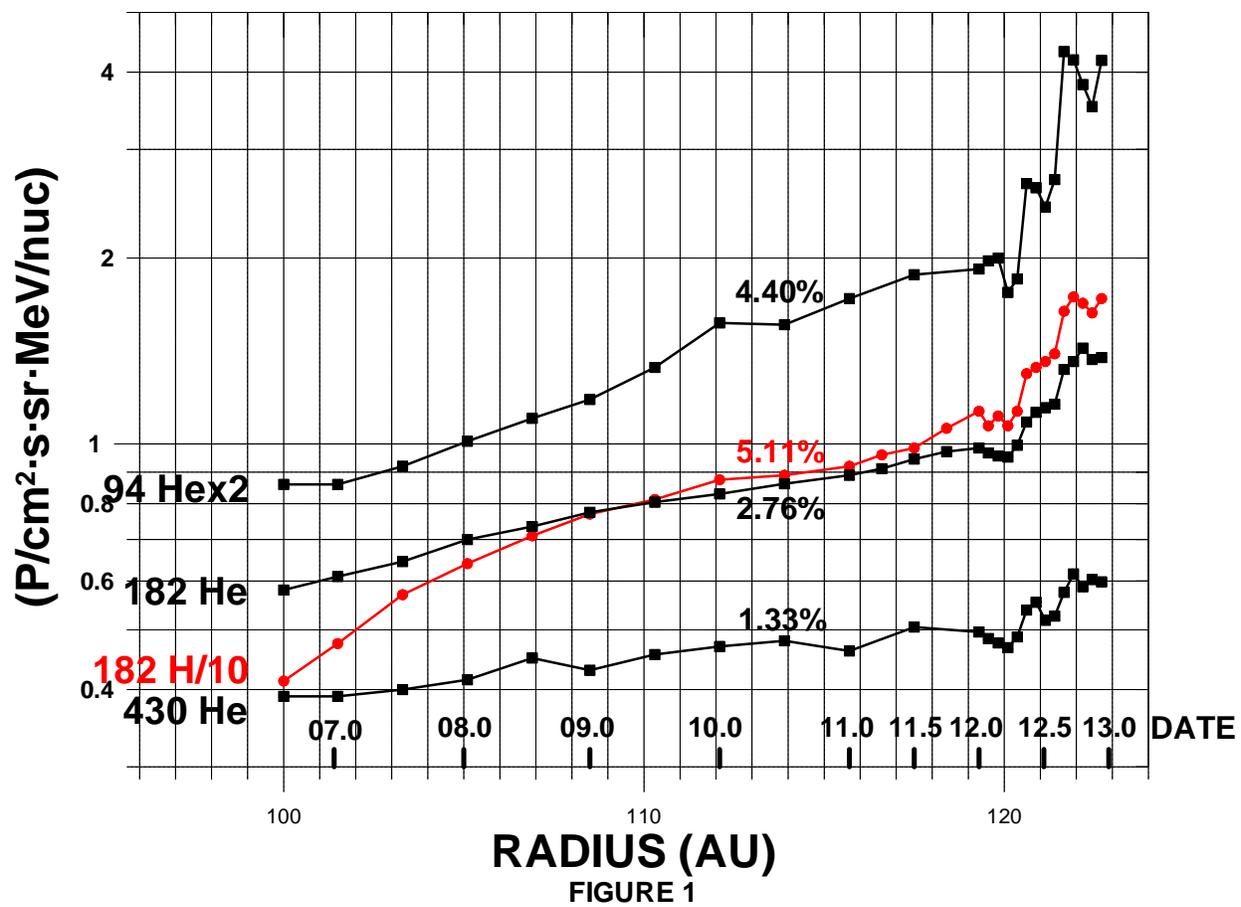

FIGURE 1

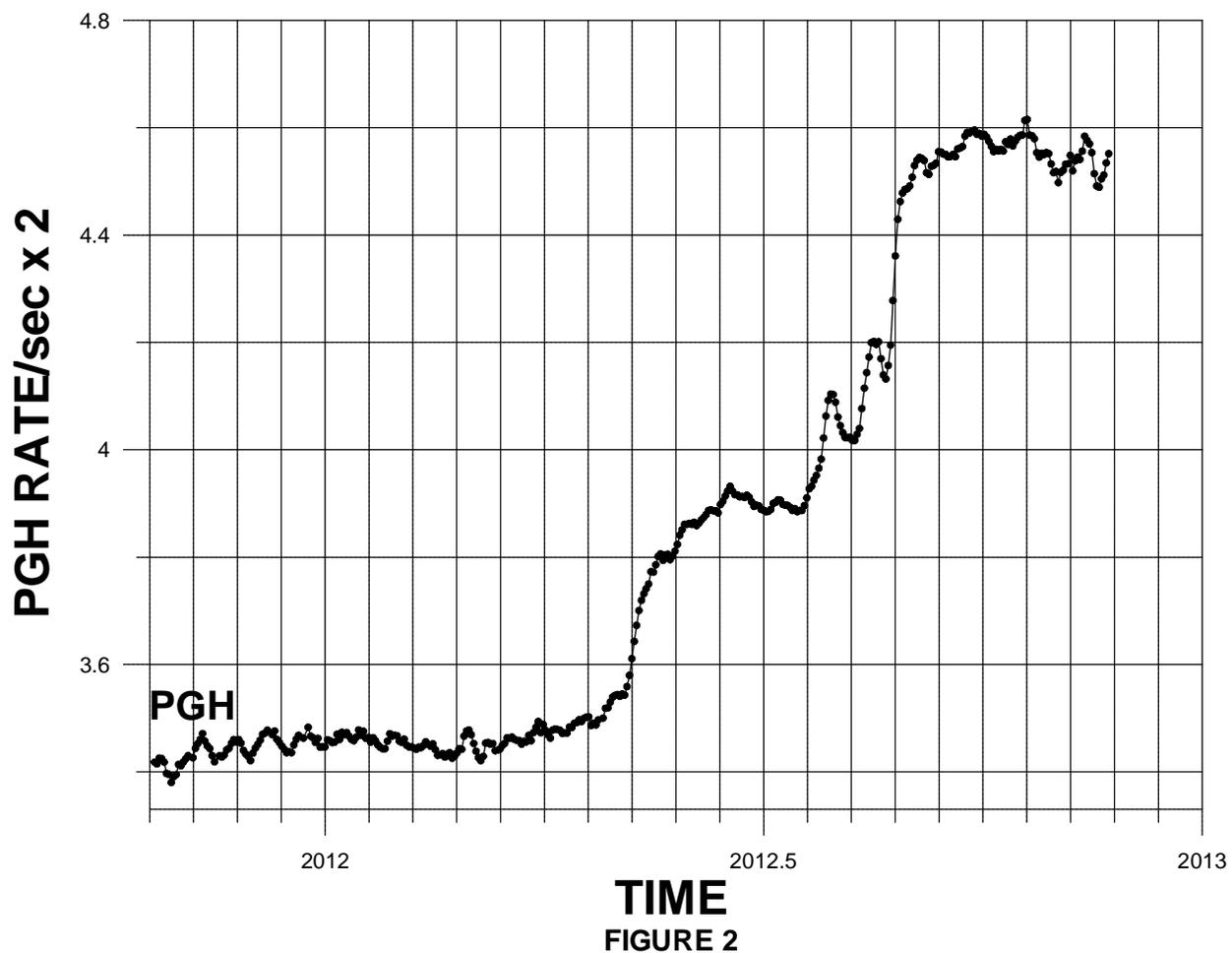

**FIGURE 2**

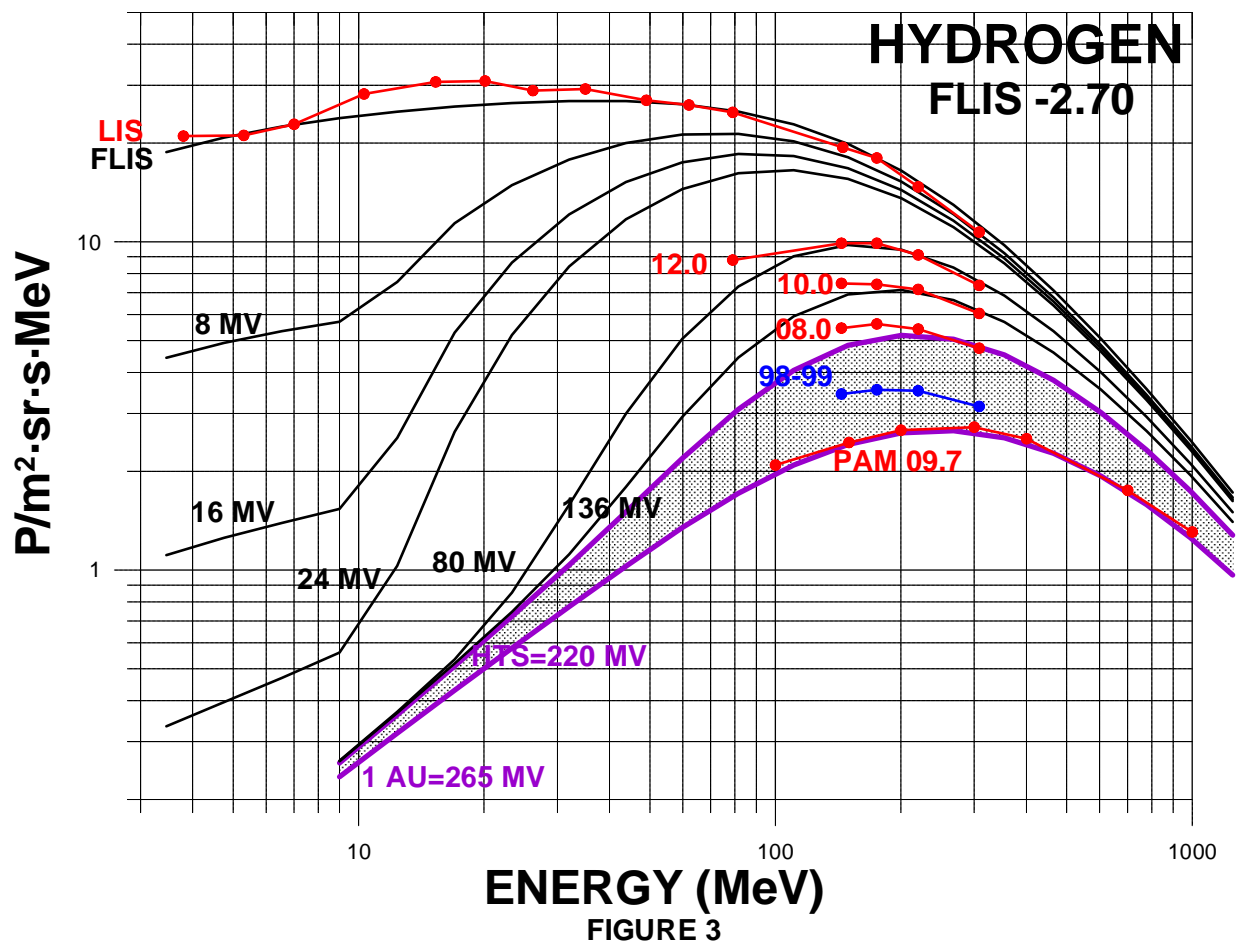

**ENERGY (MeV)**

FIGURE 3

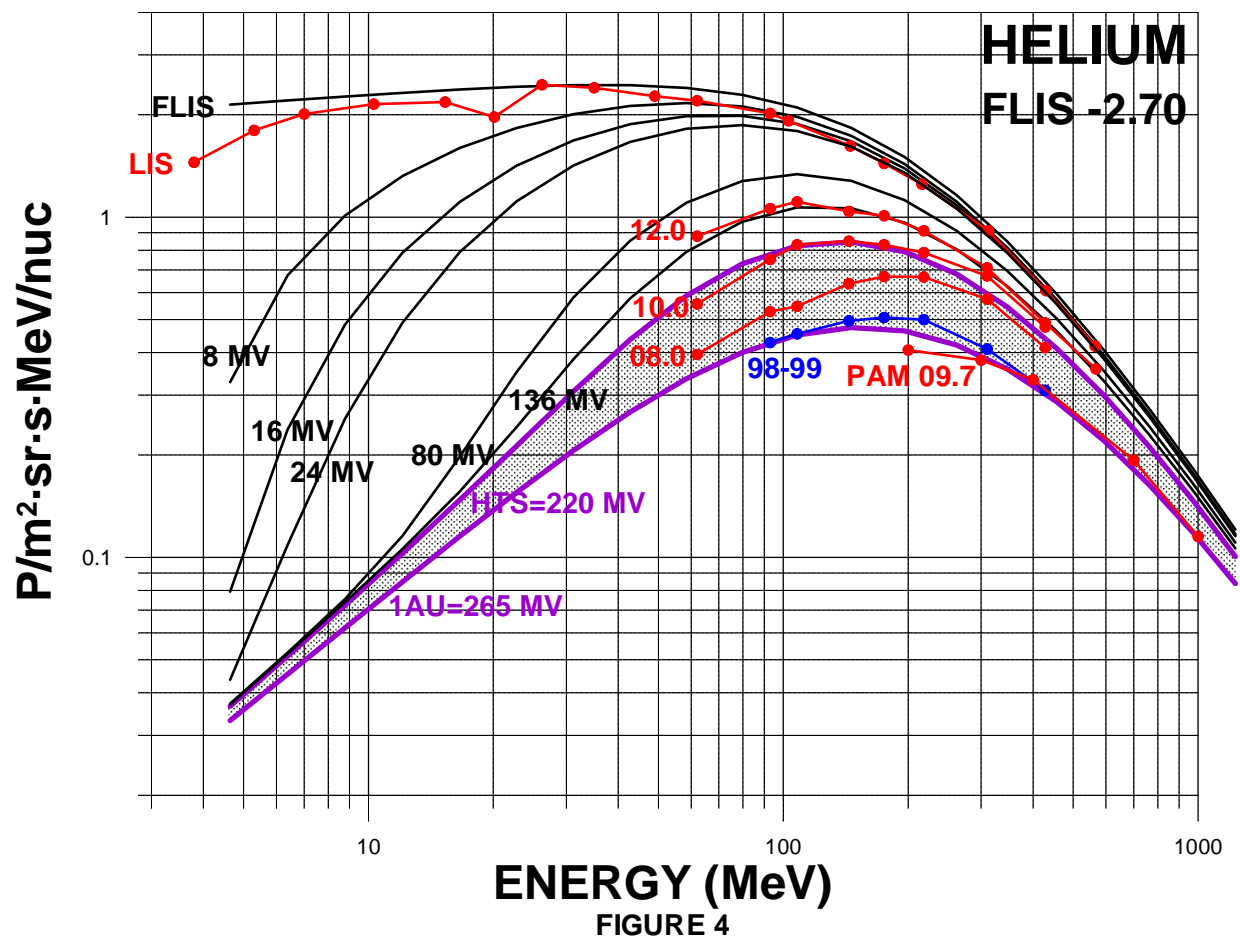

FIGURE 4

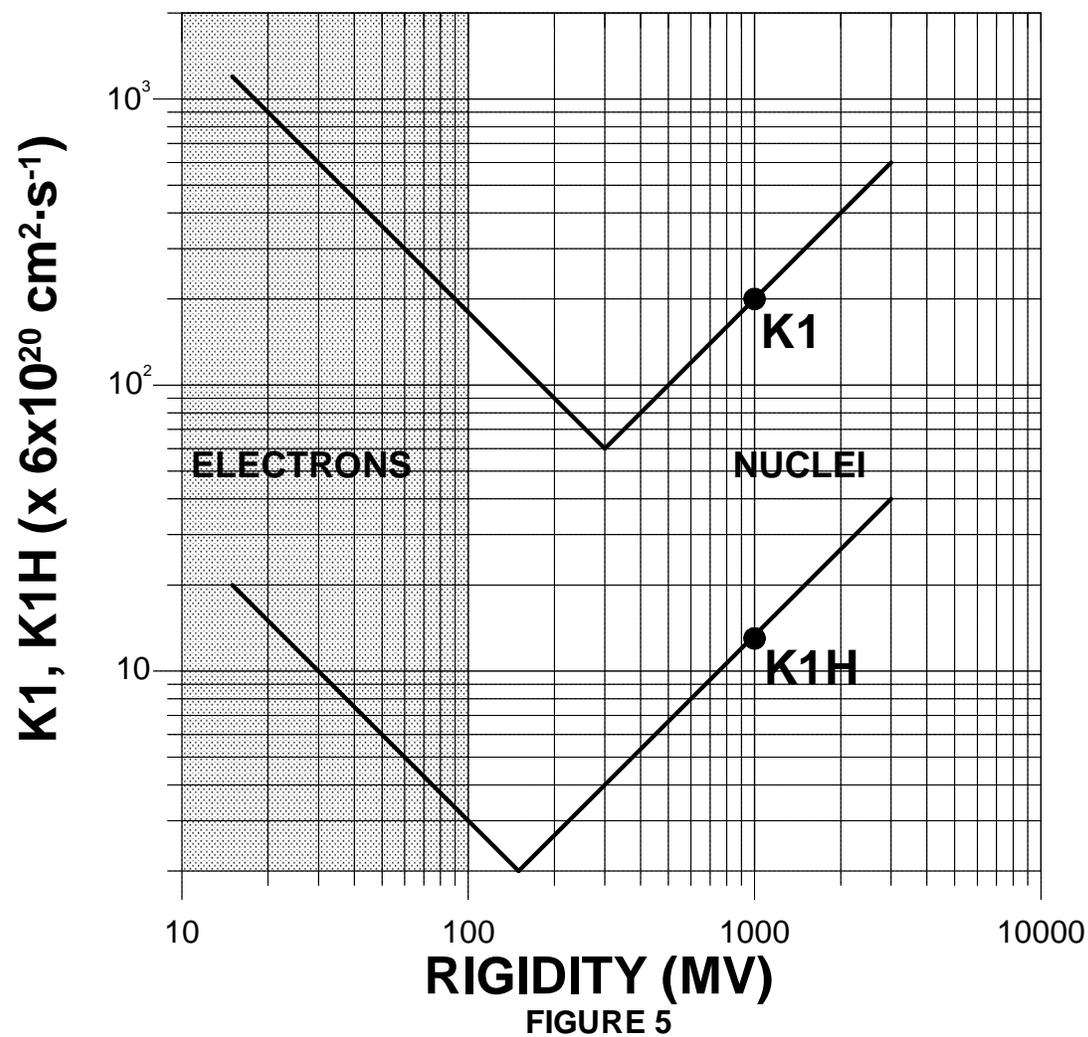

**FIGURE 5**

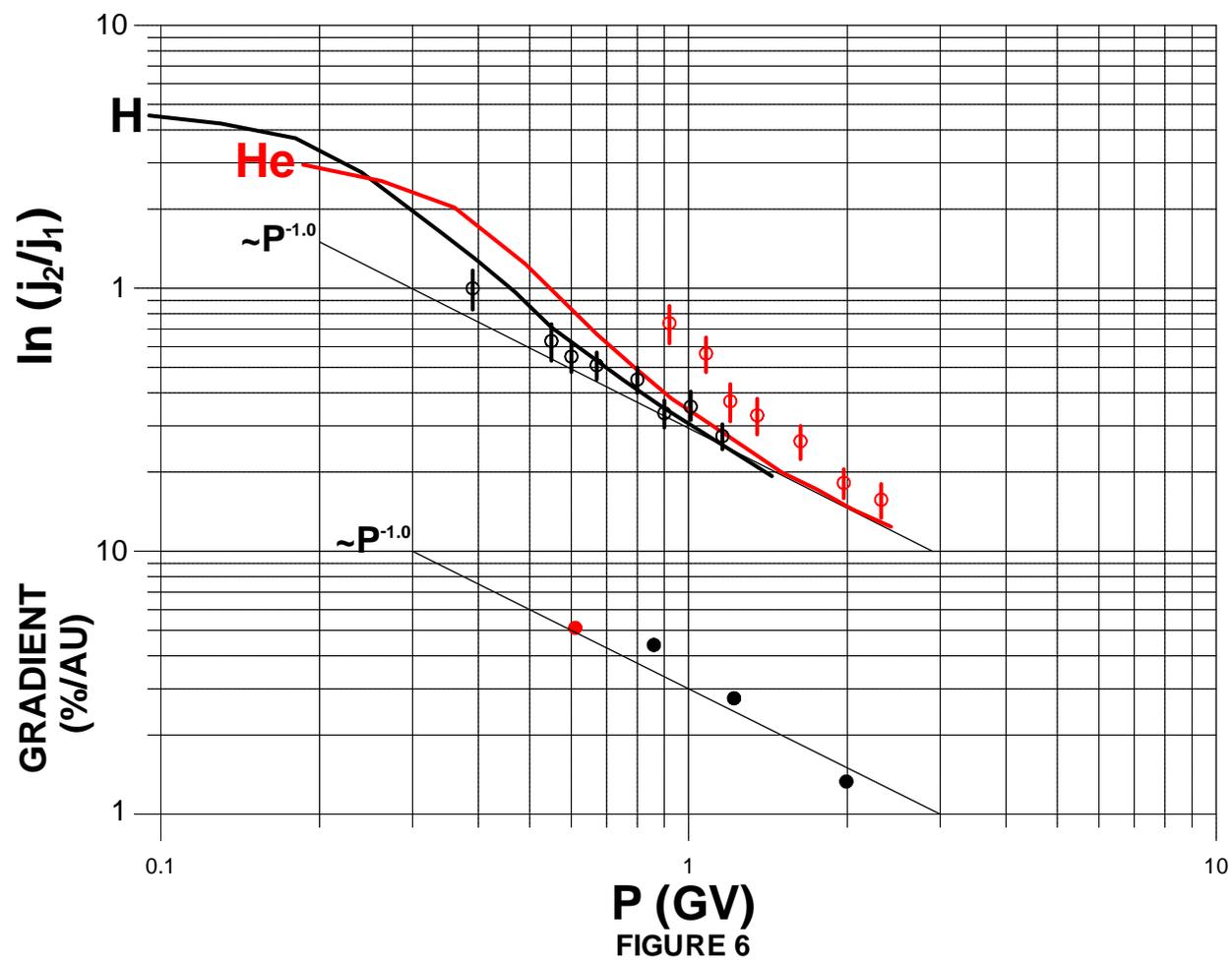

FIGURE 6